\documentclass[final]{aipproc}
\layoutstyle{6x9}
\begin{document}

\title{GRBs and the thermalization process of electron-positron plasmas}

\classification{98.70.Rz}
\keywords      {Gamma-Ray: Bursts}

\author{A.G. Aksenov}{
  address={Institute for Theoretical and Experimental Physics, B. Cheremushkinskaya, 25, 117218 Moscow, Russia}
}

\author{C.L. Bianco}{
address={ICRANet, Piazzale della Repubblica 10, 65122 Pescara, Italy and Universit\'e de Nice Sophia Antipolis, Grand Ch\^ateau, BP 2135, 28, avenue de Valrose, 06103 NICE CEDEX 2, France.}, altaddress={Dipartimento di Fisica, Universit\`a di Roma ``La Sapienza'', P.le Aldo Moro 5, 00185 Roma, Italy.}
}

\author{R. Ruffini}{
address={ICRANet, Piazzale della Repubblica 10, 65122 Pescara, Italy and Universit\'e de Nice Sophia Antipolis, Grand Ch\^ateau, BP 2135, 28, avenue de Valrose, 06103 NICE CEDEX 2, France.}, altaddress={Dipartimento di Fisica, Universit\`a di Roma ``La Sapienza'', P.le Aldo Moro 5, 00185 Roma, Italy.}
}

\author{G.V. Vereshchagin}{
address={ICRANet, Piazzale della Repubblica 10, 65122 Pescara, Italy and Universit\'e de Nice Sophia Antipolis, Grand Ch\^ateau, BP 2135, 28, avenue de Valrose, 06103 NICE CEDEX 2, France.}, altaddress={Dipartimento di Fisica, Universit\`a di Roma ``La Sapienza'', P.le Aldo Moro 5, 00185 Roma, Italy.}}

\begin{abstract}
We discuss the temporal evolution of the pair plasma created in Gamma-Ray Burst sources. A particular attention is paid to the relaxation of the plasma into thermal equilibrium. We also discuss the connection between the dynamics of expansion and the spatial geometry of the plasma. The role of the baryonic loading parameter is emphasized.
\end{abstract}

\maketitle

\section{Thermalization of the pair plasma}

\begin{figure}
	\centering
		\includegraphics[width=\hsize]{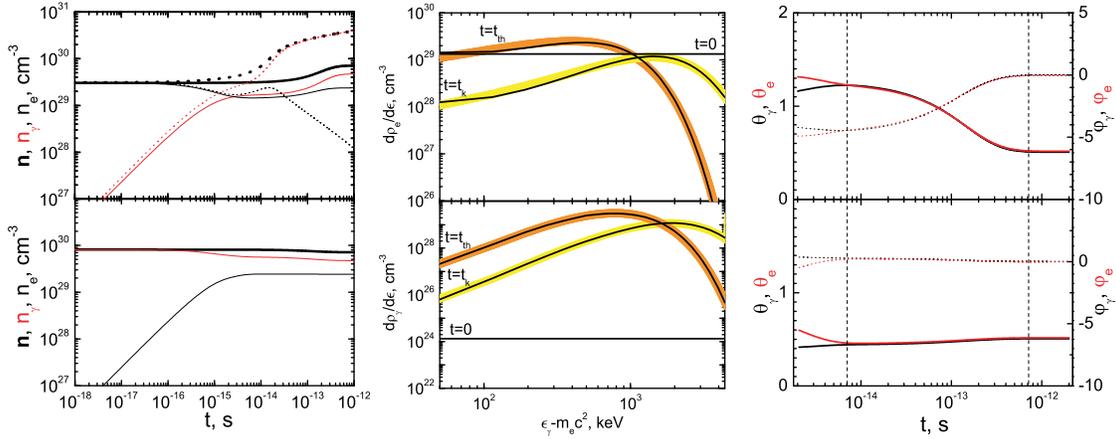}
	\label{fig1}
\caption{The density $n$ of pairs and photons (left plot, with the total density in bold), their spectra $d\rho/d\varepsilon$ (center plot) and their temperatures $\theta=kT/(m_e c^2)$ along with chemical potentials $\varphi=\phi/(m_e c^2)$ (right plot) are shown. $\varepsilon$ is the energy of particle in units of electron rest mass energy $m_ec^2$. Two different initial conditions were considered: when only pairs are present
with negligible amount of photons, and the opposite case (upper and lower
figures respectively). In both cases the pair plasma relaxes to thermal equilibrium
configuration on a timescale $t_{th}<10^{-12}$ s for our parameter range
given in Eq.(\ref{param}), i.e. much before it starts to expand on the timescale $t < 10^{-3}$ s. We also show by dashed lines (on the upper left panel) the evolution of pairs and photons concentrations when the inverse 3-body interactions are neglected. In this case we get a result similar to the one by Cavallo and Rees \cite{3}. Further details are given in Ref.~\cite{1}.}
\end{figure}

In the recent publication \cite{1} we considered the kinetic evolution of the pair plasma which is assumed to be formed in Gamma-Ray Burst (GRB) sources. Two alternative viewpoints exist in literature. In most papers, e.g. \cite{2}, it was assumed that the plasma quickly reaches thermal equilibrium, since the optical depth of the photons is much larger than one, and that it then expands adiabatically reaching a large Lorentz $\gamma$ factor. Cavallo and Rees \cite{3}, instead, suggested a number of different scenarios, assuming in all of them that the plasma cools down due to direct bremsstrahlung process until its temperature reaches $0.511$ MeV and that the electron-positron pairs disappear before starting to expand. In order to clarify this point we numerically integrated the relativistic Boltzmann equations, taking into account all 2-body and 3-body interactions between electrons, positrons and photons, assuming that the plasma is homogeneous and isotropic. We focused on the following range for the average energy $E$ per particle
\begin{equation}
	0.1\, \textrm{MeV} < E < 10\, \textrm{MeV}\,,
\label{param}
\end{equation}
which is of interest for GRBs studies \cite{4}. The results of our calculations are represented in Fig. \ref{fig1}. There, we also show how concentrations of pairs and photons evolve when inverse 3-body interactions are switched off artificially. 

\section{Fireball versus fireshell}

\begin{figure}
	\centering
		\includegraphics[width=0.5\hsize]{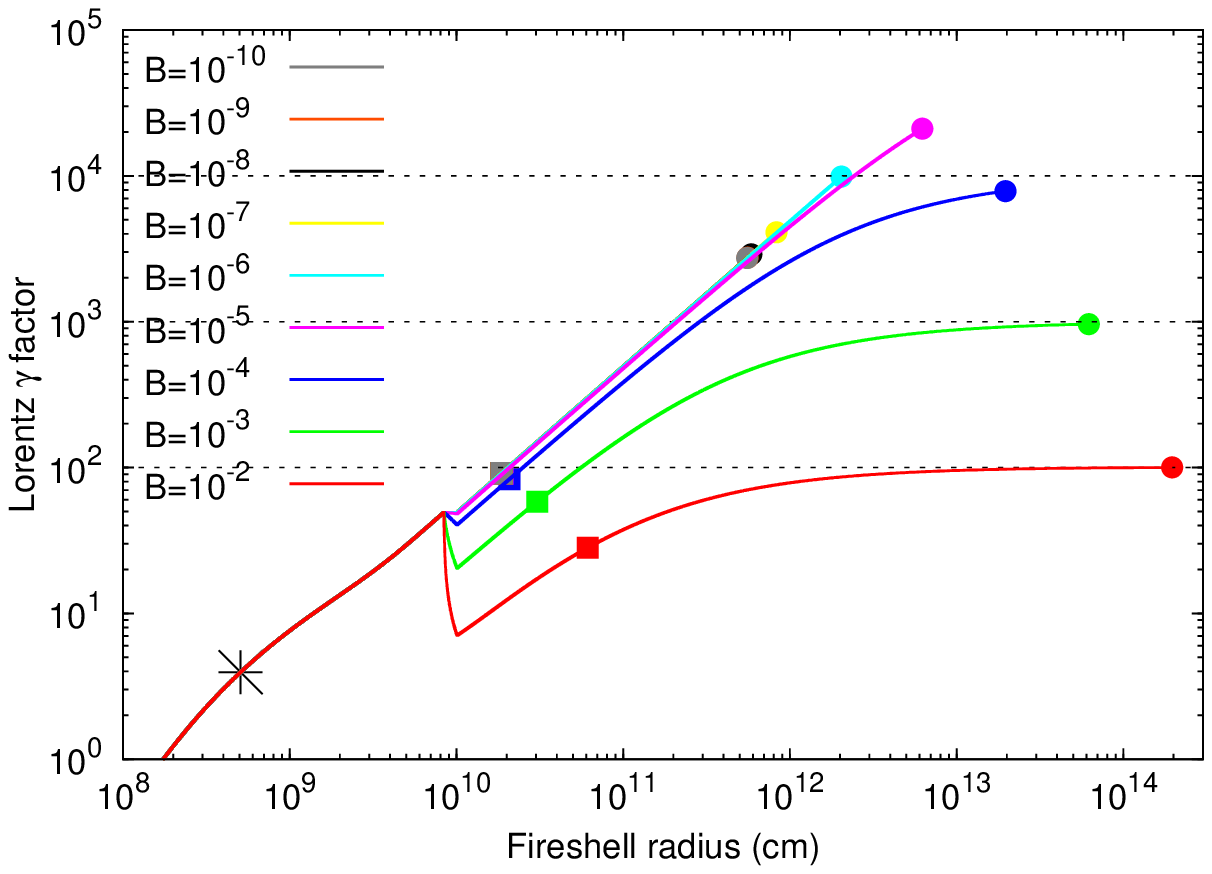}
		\includegraphics[width=0.5\hsize]{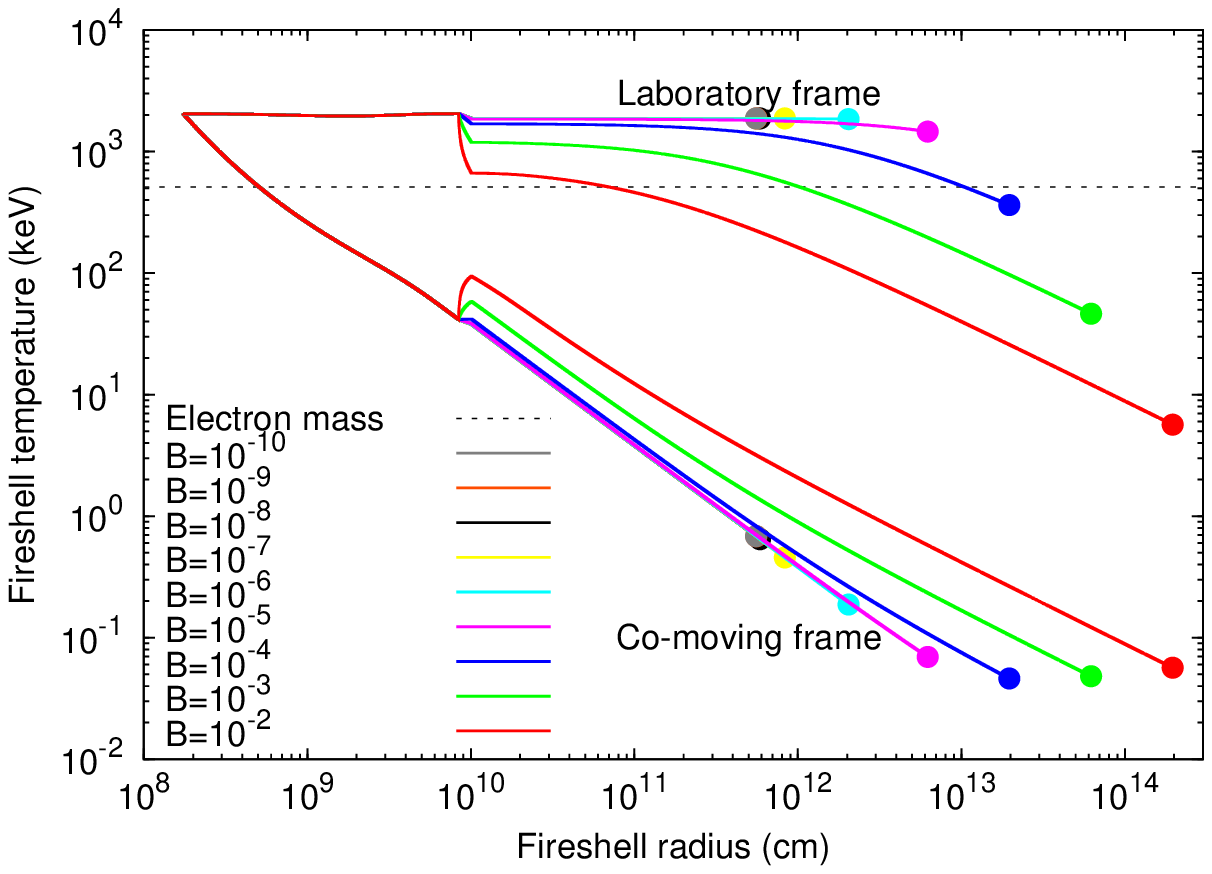}
	\caption{The fireshell Lorentz factor (left panel) and temperature (right panel) are plotted versus the radius for selected values of the baryonic loading parameter. Filled circles show the transparency point. Squares show the moment of departure from thermal distributions of electron-positron pairs. The star denotes the moment when the temperature of the fireshell equals $511$ keV. Calculations have been performed with the following initial conditions: total energy $E_{tot}=1.47\cdot10^{53}$ erg, number density of pairs $n_\pm=3.60\cdot10^{32}$ cm$^{-3}$, initial radius $R_0=1.74\cdot10^8$ cm.}
	\label{fig2}
\end{figure}

GRBs are very different from the fireball phenomenon occurring e.g. in an atomic bomb explosion. This last one is characterized by a high temperature inside, a blast wave propagating into the surrounding dense atmosphere and a characteristic self-similar motion described by the Sedov solution. The latter can be easily obtained by considerations of dimensionality. In a GRB explosion, instead, it is created a dense, hot and optically thick plasma, made by electron-positron pairs and photons, which starts to expand adiabatically being governed by equations of relativistic hydrodynamics. The surrounding medium is so rarefied that the motion of the plasma can be well represented by a relativistic expansion into vacuum. In the laboratory frame (where the black hole is at rest), the pair plasma appears to expand as a narrow shell which keeps constant its thickness due to the Lorentz contraction \cite{2}. Therefore we refer to it as a ``fireshell'' (see e.g. Refs. \cite{2007AIPC..910...55R,bianco_ita-sino} and references therein). However, we must here recall that photons emitted at the same time from the different regions of the fireshell surface reach a distant observer at different times, due to the ultrarelativistic expansion speed $\gamma\geq 10^2$ and the consequent Doppler contraction of the photons' arrival time at the detector. Therefore, a distant observer would not see the ``fireshell'' itself but the corresponding ``EquiTemporal Surfaces'' (EQTSes), i.e. the surfaces locus of points source of photons reaching a distant observer at the same time \cite{7,8,9a,9b}. During the acceleration phase, occurring because of the large radiative pressure inside, the pair plasma engulfs a certain amount of baryonic matter, parameterized by 
\begin{equation}
B = \frac{M_B c^2}{E_{tot}}\leq10^{-2},
\label{Bdef}
\end{equation}
where $M_B$ is the total mass of baryons and $E_{tot}$ is the total energy of plasma \cite{2000A&A...359..855R}. The whole system is now made by electrons, positrons, photons and protons and continues to accelerate until it becomes transparent, when the electron-positron pairs disappear and all photons escape (\cite{2,2000A&A...359..855R}, see Fig. \ref{fig2}). For simplicity, we assume a collision with a single baryonic matter shell placed between two given radii \cite{2000A&A...359..855R}.

From hydrodynamic equations it is possible to show that the thickness of the fireshell indeed remains constant under the condition $\gamma\gg 1$ \cite{5}. This condition is approached soon after the beginning of expansion since the scaling law $\gamma\propto r$ is valid for the radiation-dominated pair plasma. The above mentioned conclusion was obtained in \cite{2} by analyzing different geometries of the plasma. Notice that the broadening of the shell does not take place if the strict inequality in Eq.(\ref{Bdef}) is satisfied \cite{2000A&A...359..855R,6}.

As one can see from Fig. \ref{fig2}, the influence of baryons on the dynamics of expansion is negligible for $B < 10^{-5}$. Besides, due to the high expansion rate, the concentration of electron-positron pairs is frozen out well before the moment of transparency \cite{2000A&A...359..855R}.

\begin{figure}[ht]
	\centering
		\includegraphics[width=0.4\hsize]{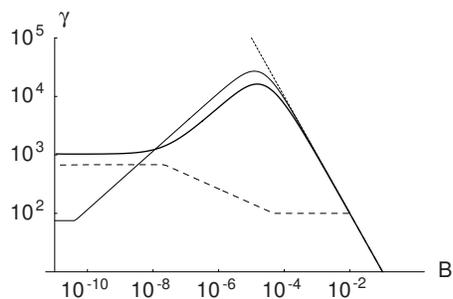}
	\caption{The fireshell Lorentz gamma factor at the transparency point is plotted versus the $B$ parameter. The thick line denotes numerical results \cite{6}, the plain line corresponds to analytical estimate from Shemi and Piran model \cite{sp}, the dotted line denotes the asymptotic value $\gamma=B^{-1}$. The dashed line shows results of Nakar, Piran and Sari \cite{nps}.}
	\label{fig3}
\end{figure}

In Fig. \ref{fig3} we show the difference between numerical \cite{2,2000A&A...359..855R} and analytical results \cite{sp,nps} for the Lorentz gamma factor reached at the transparency point \cite{6}. While the results by \cite{2} and \cite{sp} are similar for $B > 10^{-4}$, marked differences arise in the range $10^{-8} < B < 10^{-4}$, and the asymptotic values of gamma for $B \to 0$ are also different. Moreover, in the numerical treatment \cite{2000A&A...359..855R} this asymptotic behavior is reached for larger values of $B$, in disagreement with analytical expectations \cite{sp}. Thus, the acceleration of the fireshell for small $B$ is larger if one accounts for pairs dynamics, which is neglected by the analytic treatments \cite{6}.

\section{Conclusions}

We conclude that the Cavallo and Rees \emph{fireball scenarios}, a cavity filled of photons degraded to $0.5$ MeV, never apply to GRBs. They systematically neglect the occurrence of inverse 3-body interactions. If one takes into account both direct and inverse 2-body and also 3-body interactions, one obtains thermal equilibrium on a much shorter timescale than the dynamical one. Thus, the initial state of the pair plasma is described uniquely by the common temperature of pairs and photons, while their chemical potential is zero. Under these conditions the electron-positron-photon plasma is confined to a self-accelerating spherical shell, surrounding an almost empty cavity: the \emph{fireshell}.

\end{document}